\documentclass[12pt]{article}
\usepackage{epsfig,amsmath,amssymb,array,dcolumn,subfigure,rotating}

\begin{document}

\def\brho{{\mbox{\boldmath $\rho $}}}
\def\bomega{{\mbox{\boldmath $\Omega $}}}
\def\bomicron{{\mbox{\boldmath $\omega $}}}
\def\ul#1#2{\textstyle{\frac#1#2}}

\title{Correlated and Decorrelated Positional and Orientational Order in  the
Nucleosomal Core Particle Mesophases} 

\author{ V. Lorman $^1$, R. Podgornik $^{2,3,4}$  and B. \v Zek\v s $^{5,3}$
\\
$^1$ Laboratoire de Physique Mathematique et Theorique, \\
Universite Montpellier II, F-34095 Montpellier, France \\
$^2$ Department of Physics, Faculty of Mathematics and Physics, \\
University of Ljubljana, SI-1000 Ljubljana, Slovenia \\
$^3$ Department of Theoretical Physics, J.Stefan Institute, \\
SI-1000 Ljubljana,  Slovenia\\
$^4$ LPSB/NICHHD, Bld. 9  Rm. 1E-116, National Institutes of Health, \\
Bethesda, MD 20892-5626 \\
$^5$ Institute of Biophysics, Medical Faculty, \\
University of Ljubljana, SI-1000 Ljubljana, Slovenia}

\maketitle

\begin{abstract}
\small
We investigate the orientational order of transverse polarization vectors of
long columns of nucleosomal core particles and their coupling to positional
order in high density mesophases discovered recently. Inhomogeneous polar
ordering of these columns precipitates crystalization of the 2D sections with
different orientations of the transverse polarization vector on each column in
the unit cell.  We propose possible scenarios for going from the 2D hexagonal
phase into distorted lamellar and related phases observed  experimentally.
\end{abstract}

The problem of efficient genome compaction in the evolution of
eucaryotic organisms was solved on the smallest scale  {\sl via} the formation
of DNA - histone protein complexes, that make a fundamental unit of  chromatin:
the nucleosome \cite{schiessel}. A nucleosome is composed of a 147 bp long DNA
fragment, wrapped 1.75 times around the histone octamer core composed of four
different histone proteins and a DNA linker that connects two consecutive
nucleosomes \cite{luger}. Digestion of the linker DNA between different
nucleosomes along the eucaryotic genome gives rise to the {\sl nucleosome core
particle} (NCP). NCP has an approximately cylindrical shape of height and radius
$\sim$ 55 \AA, and a structural charge of $\sim$ -165. This complex is stable in
aqueous solutions from mM to 750 mM monovalent salt ionic strength
\cite{thesis}. 

At high enough concentration the solution of NCPs ceases to be isotropic.
Competing, as yet poorly understood \cite{raspaud,podgornik}, microscopic
interactions between NCPs in concentrated solutions give rise to a slew of
liquid crystalline phases of fascinating complexity \cite{MangenotJMB}.  At low
salt concentrations and sufficiently high osmotic pressure NCPs aggregate into 
columns of almost ideal cylindrical shape which contain dozens of individual
particles. These columns  then aggregate to form a lamellar phase
\cite{Leforestier}. At higher salt concentrations this phase gives way to a 2D
hexagonal columnar liquid crystalline phase leading at still higher salt
concentrations to a 3D hexagonal crystal \cite{Mangenot}.

The high asymmetry of the NCP particle, with a pronounced dyadic axis marking
the entry-exit point of the wrapped nucleosomal DNA, allows one to actually
observe the local orientational order of the NCPs, described by its in-plane
dyadic axis vector, conventionally called transverse polarization \cite{Lorman}.
In the lamellar phase the dyadic axes of individual NCPs are strongly correlated
along the column leading to a non-zero average transverse polarization vector
$\bf P$ of a column. Cryomicroscopy studies \cite{Leforestier} give strong
evidence that the dyadic axes, and consequently the vectors $\bf P$, of columns
in the two layers of a lamella are in an antiparallel direction. In addition,
the columns in the two layers are shifted to a small but constant distance in
such a way that the antiparallel dyadic axes form a periodic polar vector field
(Fig. 1). SAXS refine this picture \cite{Mangenot} by showing that there are
strong correlations between the NCP columns within a single lamella but
different lamellae appear to be completely decorrelated (Fig. 1). In this sense
the lamellar phase is decorrelated at a macroscopic scale and corresponds to
simple smectic ordering. 2D hexagonal phase is characterized, in turn, by a
positional order in the plane perpendicular to the columns, but dyadic axes of
NCPs are orientationally disordered along the columns (Fig. 1). The
corresponding average polar vector of the column is zero.

In what follows we will investigate the nature of positional and orientational
order in phases that are induced by a condensation of the periodic antiparallel
transverse polarization field modes. Basing our assumptions on experiments we
postulate that the transverse polarization order in a lamella described with
$\bf P(\brho)$, where $\brho = (x,y)$, is periodic and highly anisotropic. We
will show that the transition between the 2D colloidal crystal with disordered
dyadic axes and the smectic-like lamellar phase with the in-plane periodic polar
order can be understood by introducing a common subphase substructure for both
phases. The common subphase  is a 2D colloidal crystal with lamellar
structure and with the in-plane polar order of individual lamellae similar to
that in the actual lamellar phase. However, the correlations between the
positions of columns and the orientations of their dyadic axes from one lamella
to another make this phase crystalline and different with respect to the
smectic-like structure. The transformation from the hexagonal colloidal crystal
to the smectic lamellar phase can take place as a sequence of two second-order
transitions through the intermediate crystal lamellar phase, or as a direct
first-order transition. Order parameters of both transformations can be easily
expressed in terms of the periodic transverse polarization fields.

If the transverse polarization vector field condensation starts from a 2D
hexagonal phase then in general one would have several different choices for the
direction of the wave vector ${\bf k}$ with respect to the symmetry planes of
the 2D hexagonal phase. In the hexagonal coordinate system the wave vector of
the condensation into the crystal lamellar phase has to be described in
Kovalev's notation \cite{Kovalev} by ${\bf k}_{12} = \ul12 {\bf b}_{1}$, where
${\bf b}_1$ is one of the reciprocal basis vectors. The irreducible star of the
${\bf k}_{12}$ vector has three arms given by ${\bf k}_{1} = {\bf k}_{12}^{(1)}
= \ul12 {\bf b}_{1}$, $ {\bf k}_{2} = {\bf k}_{12}^{(2)} = \ul12 {\bf b}_{2}$
and  $ 
{\bf k}_{3} = {\bf k}_{12}^{(3)} = \ul12 \left( -{\bf b}_{1}+ {\bf
b}_{2}\right)$. The basis functions of a polar vector field with this
periodicity are given by ${\psi}_{1} = \left( 2 \hat{e}_{x} +
\hat{e}_{y}\right)~e^{i {\bf k}_{1} \cdot {\bf r}}$, ${\psi}_{2} = \left(
\hat{e}_{x} + 2 \hat{e}_{y}\right)~e^{i {\bf k}_{2} \cdot {\bf r}}$ and  $
{\psi}_{3} = \left( -\hat{e}_{x} + \hat{e}_{y}\right)~e^{i {\bf k}_{3}\cdot {\bf
r}}$. The physical realisation of the order parameter is given by the symmetry
coordinates composed of the column's polar vectors ${\bf P}_i$. Denoting by $i$
the index of the column,  in such a way that its $(x,y)$ coordinates are  
$i = 1 \longrightarrow (0,0)$, $i = 2 \longrightarrow (1,0)$, $i = 3
\longrightarrow (0,1)$ and $i = 4 \longrightarrow (1,1)$ in the 2D hexagonal
unit cell and the 2D polar vector components by $(P_x,P_y)$, one gets the
following components of the order parameter 
\begin{eqnarray}
{\eta}_{1}&=&P_{x1}\!\!-\!\!P_{x2}\!\!+\!\!P_{x3}\!\!-\!\!P_{x4}\!\!+\!\!\ul12
\left(P_{y1}\!\!-\!\!P_{y2}\!\!+\!\!P_{y3}\!\!-\!\!P_{y4}\right) \nonumber\\
{\eta}_{2}&=&\ul12\left(P_{x1}\!\!+\!\!P_{x2}\!\!-\!\!P_{x3}\!\!-\!\!P_{x4
}\right)\!\!+\!\!P_{y1}\!\!+\!\!P_{y2}\!\!-\!\!P_{y3}\!\!-\!\!P_{y4}  \nonumber\\
{\eta}_{3}&=&\ul12\left(\!\!-\!\!P_{x1}\!\!+\!\!P_{x2}\!\!+\!\!P_{x3}\!\!-\!\!P_{x4}\!\!
+\!\!P_{y1}\!\!-\!\!P_{y2}\!\!-\!\!P_{y3}\!\!+\!\!P_{y4}\right). 
\label{eq.10}
\end{eqnarray} 
The Landau free energy of this system is now given via the orientation
probability density function $\rho(x,y,z) = \rho_{0} + \sum_{i=1}^{3} \eta_{i}
\psi_{i}(x,y,z)$, in the form 
\begin{equation}
{\cal F} = \int f\left(\eta_{i}\psi_{i}(x,y,z)\right)~dxdydz = F(\eta_{i}).
\label{one}
\end{equation} 
For the sake of clarity, we have omitted in Eq. \ref{one} the terms stemming
from the large-scale elastic deformations of this order. Consequently, the free
energy of the system can be considered as a simple function of the order
parameter components. 

Let us also stress an evident duality between the polar vector carried by a NCP
column and the polar vector of its displacement in the solvent which has
important consequencies for the structure of its ordered phases. Since the
displacements ${\bf u}_i$ of NCP columns and the polar vectors ${\bf P}_i$ of
the dyadic axes span the same irreducible representation of the symmetry group
of the 2D hexagonal phase the components of the order parameter can be expressed
in the same form in terms of the columns' displacements $(u_{xi}, u_{yi})$.
Minimization of the free energy dependent on both ${\bf P}_i$ and ${\bf u}_i$
shows that they are proportional. This fact permits one to obtain also the
positional order and thus also the final structures of the phases. 

To establish the exaustive list of the phases induced by the considered
condensation mechanism we use the fact that $F(\eta_{i})$ has to be an invariant
function of the irreducible representation spanned by the periodic transverse
polarization vector field. Since the basis invariants of this represenatation
are $I_{1} = \eta_{1}^{2} + \eta_{2}^{2} + \eta_{3}^{2}$, $ I_{2} =
\eta_{1}^{2}\eta_{2}^{2} +  \eta_{2}^{2}\eta_{3}^{2} + \eta_{3}^{2}\eta_{1}^{2}$
and $ I_{3} = \eta_{1}^{2}\eta_{2}^{2}\eta_{3}^{2}$, the free energy Eq.
\ref{one} has to have the form ${\cal F} = F\left(I_{1}, I_{2}, I_{3} \right) $.
The minimization of this free energy w.r.t $\eta_{i}$ gives the equation of
state. There are six different solutions of this equation of state which we list
below in terms of $\eta_{i}$, the unit cell multiplication in the ordered state
with respect to the 2D hexagonal phase {\sl i.e.} $V_{ord}/V_{hex}$ and their
space groups.
\begin{table}[h]
\begin{tabular}{|c|c|c|}
\hline $\eta_{i}$ & $V_{ord}/V_{hex}$ & Space Group \\
\hline $\eta, 0, 0$ & 2 & $C_{2v}^{1}$ \\
\hline $\eta, \eta, \eta$ & 4 & $D_{3h}^{1}$ \\
\hline $\eta, \eta, 0$ & 4 & $D_{2h}^{1}$ \\
\hline $\eta_{1}, \eta_{2}, 0$ & 4 & $C_{2h}^{1}$ \\
\hline $\eta_{1}, \eta_{1}, \eta_{2}$ & 4 & $C_{2v}^{1}$ \\
\hline $\eta_{1}, \eta_{2}, \eta_{3}$ & 4 & $C_{S}^{1}$ \\
\hline
\end{tabular}
\caption{Different solutions of the equation of state.}
\label{table.1}
\end{table} Let us show that the first solution, {\sl i.e.} $\eta_{1} \neq 0 $,
$\eta_{2,3} = 0$ describes the crystal lamellar state. From Eq. \ref{eq.10} it
follows that in this case ${\bf P}_1 = {\bf P}_3$ and ${\bf P}_2 = {\bf P}_4$,
or in other words, that the orientations of dyadic axes carried by  columns 1
and 3 or 
columns 2 and 4 are parellel and have the same degree of correlation along the
column. Thus, due to the duality, the displacements of the columns $1$ and $3$
as well as $2$ and $4$ are also identical. This means that the periodic
transverse polarization field induces a lamellar structure, where a single
lamella is defined by the columns $1,3,1,3 ...$ for the top part and $2,4,2,4
...$ for the bottom part of a single lamella. The order parameter in this
crystalline lamellar phase induced by the periodic transverse polarization
vector ordering is given by
\begin{equation}
\eta_{1} =  (2 P_{x1} + P_{y1}) - (2 P_{x2} + P_{y2}). 
\label{two}
\end{equation}
The simplest phenomenological Landau free energy expansion of an isolated
transition from a hexagonal 2D crystal $\eta_{i} = (0,0,0)$ to a lamellar 2D
crystal with $\eta_{i} = (\eta, 0, 0)$, with basis invariants $I_{1} =
\eta^{2}$, $I_{2} = \eta^{4}$ and $I_{3} = \eta^{6}$ can be written as
\begin{equation}
{\cal F} = a_{1} I_1 + b_{1} I_2 + \dots.
\label{three}
\end{equation}
This free energy describes a second order transition between a hexagonal and a
lamellar crystalline phases, driven trivially by the parameter $a_{1}$, that can
depend on various solution conditions, e.g. the salt concentration. 
The same periodic transverse polarization vector field can induce also more
complicated transitions. Assuming now a general form of the order parameter we
can start with the following free energy {\sl ansatz}
\begin{equation}
{\cal F}_{1}(\eta_{i}) = a_{1} I_{1} + a_{2} I_{1}^{2} + b_{1} I_{2} + \dots. .
\label{four}
\end{equation} 
The phase diagram in this case is presented in Fig. 2. One can observe that in
addition to the lamellar phase, stable for $b_{1} > 0$, the diagram shows the
$(\eta, \eta, 0)$ phase, to be discussed below, stable for $b_{1} < 0$.
\begin{figure}[htb]
\begin{center}
\epsfxsize=8cm
\epsfig{file=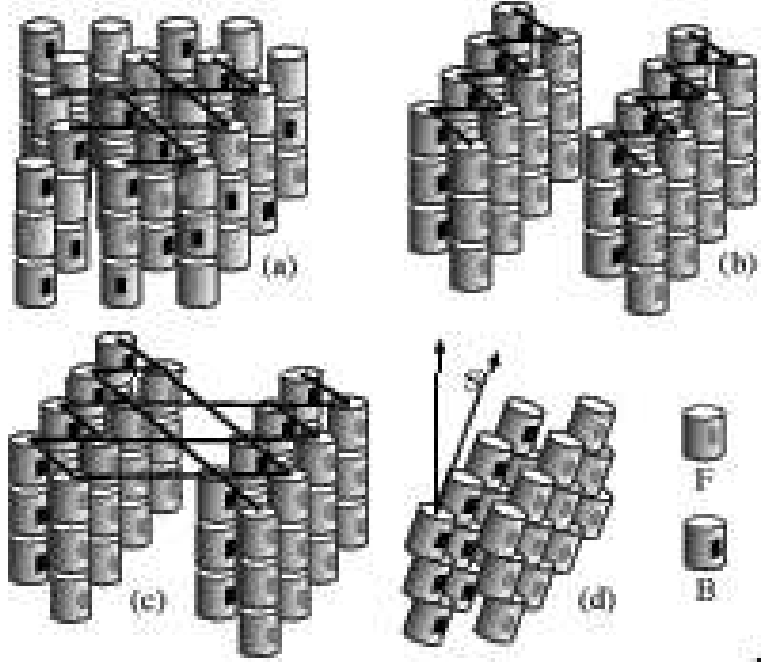, width=10cm}
\end{center}
\caption{Crystalline lamellar phase (c) as a common subphase phase of the 2D hexagonal
columnar (a) and the decorrelated smectic lamellar phase (b). Effect of
chirality on the S-axes of the NCPs (d). The red circles and black squares denote
the front (F) and the back (B) direction of the NCP dyadic axis
\cite{MangenotJMB} (lower right).}
\label{fig1}
\end{figure} The growth of the complexity of the phase diagram as more
non-linear terms are included into the Landau free energy is clearly apparent.
E.g. if one starts from the free energy
\begin{equation}
{\cal F}_{2}(\eta_{i}) = {\cal F}_{1}(\eta_{i}) + a_{3} I_{1}^{3} + c_{1} I_{3}
+ d_{12} I_{1} I_{2} + \dots.,
\label{five}
\end{equation} 
this leads to the form of the phase diagram presented on Fig. 2, where an
additional phase, this time $(\eta_{1}, \eta_{2}, 0)$, intercalates between the
$(\eta, \eta, 0)$ and the $(\eta, 0, 0)$ phases.

The structure of other phases listed in Table \ref{table.1}, induced by the
periodic ordering of the transverse polarization vector field obtained in highly
non-linear free energy models can be quite complicated. We have already analysed
the lamellar phase, beside this we also have: \\
1. $(\eta, \eta, \eta)$. The corresponding vector field is given in this phase
by the vectors of the different columns in the form :
\begin{eqnarray}
{\bf P}_1 = (P_x , 2P_y) &\qquad &  {\bf P}_2 = (0 , 0) ;  \nonumber\\
{\bf P}_3 = (P_x , -P_y) &\qquad &  {\bf P}_4 = (-2P_x , -P_y) 
\end{eqnarray}
The vectors of orientation (and, consequently, the displacements) of the columns
are expressed in terms of one polar vector $(P_x , P_y)$. Note that, because of
the specific interaction between the columns in this phase, column 2 is
frustrated : its average polar vector is zero. Consequently, its displacement
with respect to the position in the 2D hexagonal phase is zero and the dyadic
axes of individual NCPs are disordered along this column. All three components
of the order parameter are equal to the following combination of ${\bf P}_i$
vectors : $\eta = P_{x1} + 2P_{y1} +  P_{x3} - P_{y3} - 2P_{x4} - P_{y4}$.
Analyzing the displacement field in this phase, see Fig. 2, it is obvious that
the columns of a hexagonal phase are now displaced to make an inverted hexagonal
phase, observed also in experiments  \cite{MangenotJMB},
whose vertices are given by the columns $3'$, $2$, $1'$, $2$, $4'$ and $2$.\\
2. $(\eta, \eta, 0)$. The corresponding vector displacement fields in this phase
are given by
\begin{eqnarray}
{\bf P}_1 = \ul32(P_x , P_y) &\qquad &  {\bf P}_2 = \ul12 (-P_x , P_y)  
\nonumber\\
{\bf P}_3 = \ul12(P_x , -P_y) & \qquad &  {\bf P}_4 = -\ul32(P_x , P_y) 
\end{eqnarray}
Both non-zero components of the order parameter in this phase are equal to $\eta
= \ul32(P_{x1} + P_{y1}) + \ul12(-P_{x2} + P_{y2}) + \ul12(P_{x3} - P_{y3}) -
\ul32(P_{x4} + P_{y4})$. The columns of a hexagonal phase undergoing the
orientational and displacement fields, see Fig. 2, form now an orthorhombic
phase with a large unit cell. It can be understood as a succession of two series
of columns undergoing 1D displacement waves : in the first series the wave is
transeverse and in the second one it is longitudinal. Note that the point
symmetry of this structure shows only slight deviation from the hexagonal
symmetry. \\
3. Along the same line the vector fields and the structures of the low-symmetry
phases $(\eta_{1}, \eta_{2}, 0)$,  $(\eta_{1}, \eta_{1}, \eta_{2})$ and
$(\eta_{1}, \eta_{2}, \eta_{3})$ can be obtained.

We have shown how a 2D ordering of a transverse vector polarization field can
induce a transition into a lamellar crystalline phase. The experimentally
observed lamellar phase is however not of this type and is basically equivalent
to a smectic phase. The difference is subtle and referrs to the fact that in
SAXS in addition to the smectic peaks only reflections with two non-zero indices
from the NCPs correlated along the column and within a single lamella are
present. No reflection with three non-zero indices which can evidence the
correlation between different lamellae is observed. The lamellae are thus
decorrelated and the 3D long range order of the transverse polarization vector
is destroyed. This means that the macroscopic symmetry of this phase is higher
than that of the crystal lamellar phase with $\eta_{i} = (\eta, 0, 0)$ and
equivalent to the symmetry of a usual lamellar smectic phase ($G=D_{\infty h}\,
\otimes \, T_{d}^{z}$). One can speculate that this decorrelation is related to
the difference in correlation lengths of microscopic interactions which insure
the stability of NCP mesophases on the one hand and their colloidal character on
the other hand. Indeed, the displacements of columns in a 2D colloidal crystal
are much greater than typical values of atomic shifts in a close-packed
hexagonal crystal. Resulting distance  $d_L$ between the lamellae in our case
can become greater than the correlation length of the interaction between the
NCPs but still smaller then the correlation length of the smectic density wave
$\xi_{\rho(z)}$.
\begin{figure}[htb]
\begin{center}
\epsfxsize=8cm
\epsfig{file=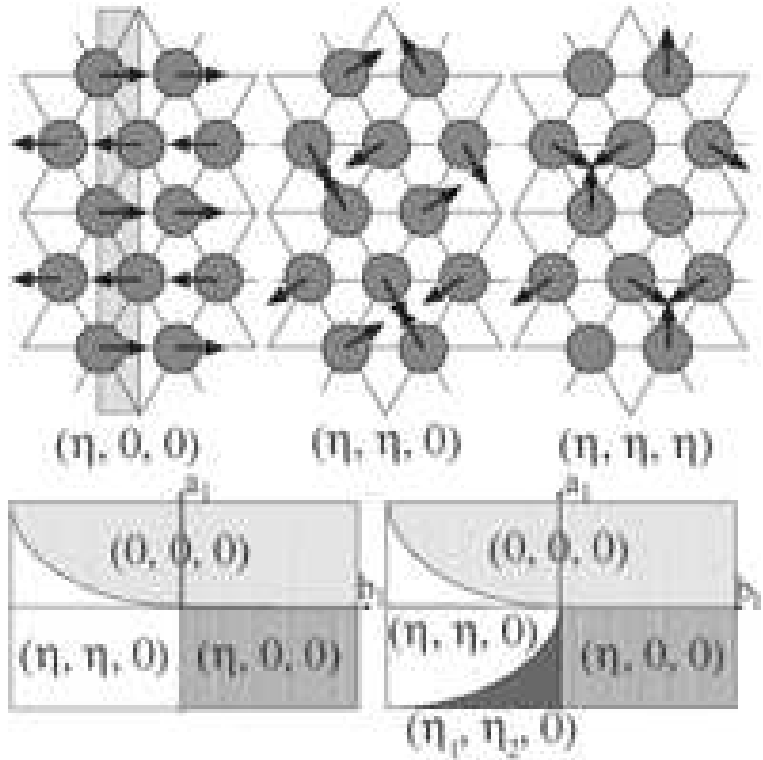, width=10cm}
\end{center}
\caption{Displacement fields for the transition from the hexagonal to lamellar
$(\eta, 0, 0)$ (columns forming a single lamella are enclosed within a gray
rectangle), orthorhombic $(\eta, \eta, 0)$ and inverted hexagonal $(\eta, \eta,
\eta)$ phases. Below, the phase digram stemming from Eqs. \ref{four},
\ref{five}.}
\label{fig2}
\end{figure} Since the symmetry of the crystal lamellar phase is lower than that
of the actual lamellar smectic phase, this decorrelation transition must be
accompanied by vanishing of a physical quantity responsible for this structural
change. In other words the order parameter of the lamellar correlation
transition can be defined. Group-subgroup relationship between these two lamellar
phases permits one to show that the physical quantity vanishing through
decorrelation transition is the macroscopic transverse polarization vector wave
with the wave vector parallel to the smectic layer and the polarization directed
perpendicular to the single lamella (1D transverse vector wave). Condensation of
this wave leads directly from the smectic-like decorrelated structure to a 
correlated crystal lamellar phase of the $\eta_{i} = (\eta, 0, 0)$ type, with
periodic distribution of transverse polarization in columns in each layer. The
order parameter of the transition is given by the first harmonic of the wave
{\sl i.e.} $ u^{z}(k) \hat{e}_{z} exp{(iky)} + u^{z}(-k) \hat{e}_{z}
exp{(-iky)}$, where $u^{z}(k)$ is the amplitude of this harmonic  and $k$ is the
wave vector perpendicular to the direction of the columns within a lamella, with
the magnitude $k = \frac{2\pi}{a}$, where $a$ is the distance betweem
neighboring columns within a single lamella.

These polarization waves exist localy for each lamella, but the distribution of
the origins of these waves in each lamella is random for the whole sample and
consequently the average macroscopic value of the wave is zero in the
decorrelated phase (Fig. 1). Similar scenario but in a different context was
first proposed by P. Bak \cite{Bak}. The order-disorder character of the
transition and the wave nature of the order parameter were clarified. It was
also shown that the wave vector of the macroscopic order parameter for the
transitions of this type is equal to the wave vector of the local waves (wave
vector of an atomic chain in \cite{Bak} or of the polar vector wave in an
individual lamella in our case). The transition from the decorrelated smectic
phase into the correlated $\eta_{i} = (\eta, 0, 0)$ type phase is described
trivially via the Lifshitz free energy \cite{Landau} $ {\cal F} = \alpha_{1}
\vert u^{z}(k)\vert^{2} + \alpha_{2} \vert u^{z}(k)\vert^{4} + \dots $ and is
thus of second order.

In this correlated-decorrelated transition scenario the crystalline lamellar
phase is a common subphase of both the hexagonal crystalline phase as well
as the smectic lamellar phase. The hexagonal crystal $\longrightarrow$ crystal
lamellar phase transition is driven by condensation of the 2D periodic
transverse polarization field, whereas the smectic lamellar  $\longrightarrow$
crystal lamellar phase transition is driven by the condensation of the 1D polar
vector wave  and is characterised by the correlation between the transverse
polarization vector distributions in each individual lamella.

Let us finally discuss the influence of chirality on the state of NCP columns.
Each NCP particle being chiral their aggregation results in a chiral column with
the rotational symmetry of a cylinder without neither mirror planes nor
inversion, and the periodicity approximately equal to the individual NCP height.
According to the Curie principle ordering of the polar vector ${\bf P}_i$ along
a chiral column leads to a simultaneous ordering of the axial vector ${\bf
t}_i$, since both ${\bf P}_i$ and ${\bf t}_i$ span the same representation of
the column symmetry group. In a chiral periodic aggregate the axial vector has
an evident physical interpretation. Non-zero average value of ${\bf t}_i$
corresponds to the correlated tilt of individual NCPs with respect to the column
axis (Fig. 1). In the 2D hexagonal phase the NCP columns carry zero average
polar vector, then in a column the tilt of the particles is also zero : the axes
of individual particles (usually called S-axes) are parallel to the column axis.
On the contrary, in the lamellar phase the ordering of the polar dyadic axes
induces a tilt of the NCPs S-axes. This simple consequence of the NCP
chirality can explain peculiarities of the SAXS spectrum observed in
\cite{Mangenot}.

One of the authors (R.P.) would like to thank F. Livolant and her collaborators
for their hospitality while being a Chercheur Associ\' e fellow of C.N.R.S at
Orsay in 2002.

\end{document}